\documentstyle[12pt]{article}
\title{A Symmetry Group of a Thue-Morse Quasicrystal}
\author{Jean-Pierre Gazeau\footnotemark[1]
\\ Laboratoire de Physique Th\'{e}orique
de la Mati\`{e}re Condens\'{e}e\\ Universit\'{e} Paris 7 - Denis Diderot \\
2, place Jussieu\\
F-75251 Paris Cedex 05\\ France \\ [10mm]
Jacek Mi\c{e}kisz\footnotemark[2] 
\\ Institute of Applied Mathematics and Mechanics \\
University of Warsaw\\ ul. Banacha 2, 02-097 Warsaw \\ Poland}
\begin{document}
\maketitle
\baselineskip=14pt
PACS numbers: 05.20.-y, 05.50.+q, 61.44.Br
\vspace{10mm}

* E-mail address: gazeau@ccr.jussieu.fr

$\dag$  E-mail address: miekisz@mimuw.edu.pl
\eject
\noindent {\bf Abstract.} We present a method of coding general
self-similar structures. In particular, we construct a symmetry group
of a one-dimensional Thue-Morse quasicrystal, i.e., 
of a nonperiodic ground state of a certain translation-invariant, 
exponentially decaying interaction. 
\vspace{10mm}
 
\newtheorem{prop}{Proposition}
\newtheorem{theo}{Theorem}
\noindent A symmetry group of a three-dimensional crystal
consists of lattice translations, rotations, and reflexions. Starting
from any point of a crystal, we can reach any other point, successively
applying different elements of the symmetry group of the crystal.
It was shown recently \cite{gaz1,gaz2} that certain 
one-dimensional quasicrystals
can be build by successive applications, on one of its points, of elements
of certain discrete affine semigroups. Here we describe a general method,
based on ideas contained in \cite{gru,conn}, of representing self-similar
structures by one-sided sequences of two symbols. In particular, we 
construct a symmetry group of a Thue-Morse quasicrystal, i.e., of a 
nonperiodic ground state of a certain one-dimensional classical 
lattice-gas model.

In one-dimensional classical lattice-gas models, every site 
of the lattice ${\bf Z}$ (the set of all integers) 
can be occupied by a particle or be empty. 
Configurations of such models are therefore 
elements of $\Omega=\{0,1\}^{{\bf Z}}$,
where $1$ denotes the presence and $0$ the absence of a particle
at any given lattice site. By $X(i)$ we denote the projection of $X$
to a lattice site $i \in {\bf Z}$.
\vspace{3mm}

\noindent {\bf The Thue-Morse Example.}
We use the Thue-Morse substitution rule to construct 
a configuration in $\Omega$. We put $1$ at the origin 
and perform alternatively to the right and to the left
the following substitution: $1 \rightarrow 10, 0 \rightarrow 01$.
After the first right substitution we get $10$, which 
is the configuration on $[0,1]$, then the substitution 
to the left gives us $0110$ on $[-2,1]$ (the substitution 
to the left means that we replace $1$ by $01$ and $0$
by $10$), then we get $01101001$ on $[-2,5]$ and so on.
The effect of performing the substitution to the right or to the left 
is the same as taking a sequence of symbols already obtained, 
changing every $0$ to $1$ and every $1$ to $0$ and then placing 
the new sequence either right to or left to the previous sequence.

In this manner we obtain a nonperiodic configuration $X_{TM} \in \Omega$.
Notice that if all substitutions were performed to the right, we would
get the well-known one-sided Thue-Morse configuration: $1001011001101001...$
\cite{pru,th,mor}  

Let $T$ be a translation operator, i.e., $T:\Omega \rightarrow \Omega$,
$T(X)(i)=X(i-1)$, $X \in \Omega$. Let $G_{TM}$ be a closure (in the product 
topology of the discrete topologies on $\{0,1\}$) of the orbit
of $X_{TM}$ by translations, i.e., $G_{TM}=\{T^{n}(X_{TM}), n \geq 0\}^{cl}$.
We call elements of $G_{TM}$ two-sided Thue-Morse configurations.
It can be shown that $G_{TM}$ supports exactly one translation-invariant
probability measure $\mu_{TM}$ on $\Omega$ \cite{kak,kea1}. Such measure
is called uniquely ergodic and can be obtained as the limit of averaging over
$X_{TM}$ and its translates:
$\mu_{TM}=lim_{n \rightarrow \infty}(1/n)
\sum_{i=1}^{n}\delta(T^{i}(X_{TM}))$,  
where $\delta(T^{i}(X_{TM}))$ is the probability measure
assigning probability $1$ to $T^{i}(X_{TM})$. 
It means that all two-sided Thue-Morse configurations
look locally identical - any local pattern of particles appears
in all of them with the same density (defined uniformly in space).
It is also said that such configurations belong to the same
isomorphism class. 

The Thue-Morse measure is the unique ground state, i.e.,
a measure supported by configurations with the minimal energy density,
of certain exponentially decaying, translation-invariant, four-body 
interaction \cite{ja1,ja2,emz}. Multilayer Thue-Morse superlattice
heterostructures were recently made by means of the Molecular Beam Epitaxy
and were investigated by Raman scattering \cite{mbe1} and high resolution
X-ray diffraction \cite{mbe2}. Nonperiodic order present in Thue-Morse 
configurations (and in Fibonacci configurations defined below)
was investigated in \cite{ja3,ja4,horn}.
\vspace {3mm}

\noindent {\bf The Fibonacci Example.} We repeat the above procedure
using the Fibonacci substitution:
$0 \rightarrow 01, 1 \rightarrow 0$. We put $0$ at the origin and apply 
alternatively the above substitution to the right and to the left. We obtain 
the configuration $X_{F} \in \Omega$. If all substitutions were performed 
to the right we would get the right one-sided Fibonacci 
configuration $01001010...$ Denote by $G_{F}$ the closure 
of the orbit of $X_{F}$ by translations. The elements of $G_{F}$ are called
two-sided Fibonacci configurations.
$G_{F}$ supports the uniquely ergodic measure
$\mu_{F}$ which is the unique ground state of any exponentially decaying,
strictly convex, repulsive interaction 
and a chemical potential which fixes the density
of particles to be equal to the square of the golden ratio 
$(2/(1+\sqrt{5}))^{2}$
\cite{bak,aub,ja5}. 
\vspace{3mm}

We will now discuss a concept of self-similarity.
Let $X \in \Omega$. Let us assume that there are two types of 
finite configurations, denote them by $s_{0}, s_{1}$, such that
we can group all symbols of $X$ into successive local configurations 
of these types. We then construct $Y_{X} \in \Omega$ in the following way.
If $0 \in {\bf Z}$ belongs to the support of a local configuration 
of the $s_{j}$ type, then $Y_{X}(0)=j$, $j=0, 1$.
Now, let $s^{i} \in \{s_{0}, s_{1}\}, i=1, 2,...$ be successive types of
local configurations right to the origin and 
$s^{i} \in \{s_{0}, s_{1}\}, i=-1, -2,...$ be
successive types of configurations left to the origin. If $s^{i}=s_{j}$,
then we define $Y_{X}(i)=j$. 
\vspace{2mm}

\noindent {\bf Definition:} A set of configurations, $G \subset \Omega$,
is {\bf self-similar} if for some choice of $s_{0}, s_{1}$,
for every $X \in G$, $Y_{X} \in G$. Then the grouping of symbols of $X$
into local configurations of the type $s_{0}$ or $s_{1}$ 
is also called self-similar.
\vspace{3mm}

\noindent The following proposition has been proven in \cite{sen}.
\begin{prop}
If a self-similar grouping of symbols, into given local configurations
$s_{0},s_{1}$, of a configuration $X \in \Omega$ from a self-similar set 
is unique, then $X$ is nonperiodic.
\end{prop}
{\bf Proof by contraposition}: Let us assume that $X$ has period $p$.
We repeat the process of grouping symbols until 
$p < \max \{|c_{0}|, |c_{1}|\}$,
where $c_{0},c_{1}$ are local configurations of $X$  such that after
all successive groupings they are respectively of $s_{0}$ and $s_{1}$ type
and $|c_{0}|,|c_{1}|$ are lengths of their supports. Now we translate 
$X$ by $p$ lattice units. $X$ does not change.
However, $c_{0}$ and $c_{1}$ overlap with their translates. 
The translation produces therefore a different grouping.
\vspace{3mm}

\noindent {\bf The Thue-Morse Example.} The set $G_{TM}$ 
of two-sided Thue-Morse configurations is self-similar 
and a corresponding self-similar grouping is the following: 

\noindent $s_{0}=01 \rightarrow 0, s_{1}=10 \rightarrow 1$. 
\vspace{3mm}

\noindent {\bf The Fibonacci Example.} The set $G_{F}$ 
of two-sided Fibonacci configurations is self-similar 
and a corresponding self-similar grouping is the following:

\noindent $s_{0}=01 \rightarrow 0, s_{1}=0 \rightarrow 1.$ 
\vspace{3mm} 

\noindent We present now a method of coding self-similar sequences.

Let $G \subset \Omega$ be a self-similar set and let $X \in G$.
If $X(0)=1$, then we set $C_{X}(-1)=1$; if $X(0)=0$, then $C_{X}(-1)=0$.
If $0 \in {\bf Z}$ belongs to the support of $s_{j}$, 
then $C_{X}(0)=j$. Now we group the 
symbols of $X$ and construct $Y_{X}$. If $0$ belongs to the support of
$s_{j}$ of $Y_{X}$, then $C_{X}(1)=j$. 
We group the symbols of $Y_{X}$, construct $Y_{Y_{X}}$
and obtain $C_{X}(2)$. We continue this procedure infinitely many times
and obtain a sequence $C_{X}(i), -1 \leq i < \infty$.   
$C_{X}$ can be seen as an element of a direct product 
${\cal W}= \otimes_{i=-1}^{\infty} Z_{2}$,
where $Z_{2}$ is the group of two elements, $0$ and $1$, with the addition
modulo $2$ as a group action. 
\vspace{3mm}

\noindent {\bf The Fibonacci Example.} The Fibonacci 
configurations are represented
by elements of ${\cal W}_{F}=\{W \in {\cal W}, W(i)W(i+1)=0$, for every
$i \geq -1\}$. This restriction on $W's$ is present in a 
corresponding representation of Penrose tilings \cite{gru,conn}.

If $C_{X} \in {\cal W}_{F}$ has both infinitely many 0's and 1's,
then $X$ is a two-sided Fibonacci configuration, 
i.e., an element of $G_{F}$. Otherwise,
$X$ is a one-sided Fibonacci configuration. For example, 
$(0010101010...)$ represents $X_{F}$
and $(0000000...)$ represents the right one-sided Fibonacci 
configuration.
\vspace{3mm}

\noindent {\bf The Thue-Morse Example.} It is easy to see that every
element of ${\cal W}$ represents either a two-sided Thue-Morse configuration, 
i.e., an element of $G_{TM}$ or a one-sided Thue-Morse configuration. 
For example, $(11001100...)$
represents $X_{TM}$, $(111111...)$ represents
the right one-sided Thue-Morse configuration (with $1$ at the origin)
and $(101010...)$ represents the left one-sided Thue-Morse sequence 
$...01101001$ (with $1$ at the origin) obtained
by successive applications of the Thue-Morse substitution to the left.
Let us notice that the representation of the last configuration
has infinitely many $0$'s and $1$'s.
However, we would like to represent one-sided configurations by sequences
with either finitely many $0's$ or finitely many $1$'s. To achieve this,
we represent Thue-Morse configurations in another way. 

Observe that every Thue-Morse configuration can be obtained 
by successive applications of the Thue-Morse substitution either to the right
or to the left. If $X(0)=1$, then let $C'_{X}(-1)=1$; if $X(0)=0$, then
let $C'(-1)=0$. We put $C'_{X}(i)=1$ if the $i$-th substitution was performed
to the right and $C'_{X}(i)=0$ if the $i$-th substitution was performed 
to the left. Such $C'_{X}$ is again an element of ${\cal W}$. 

Now, the right one-sided Thue-Morse configuration is still represented by
$(111111...)$ but the left one is represented by $(100000...)$ and $X_{TM}$
by $(11010101...)$. It is easy to see that 
$C'_{X}(i)=(C_{X}(i-1)+C_{X}(i)+1)$ mod 2. $X \in G_{TM}$ if and only if
$C'_{X}$ has both infinitely many $0$'s and $1$'s.    

Obviously, for some of the elements of $\cal W$, corresponding Thue-Morse
configurations are related by translations, reflexions or the transformation
changing every $0$ to $1$ and every $1$ to $0$. We say that such 
Thue-Morse configurations are equivalent. Let $\cal U \subset \cal W$ be
a subgroup of $\cal W$ consisting of sequences with finitely many $1$'s
or finitely $0$'s. We have the following theorem.
\begin{theo}
Equivalent classes of Thue-Morse configurations are represented by elements 
of the coset group $\cal W/\cal U$.
\end{theo}
{\bf Proof:}  
\vspace{1mm}

\noindent {\bf a)} Let $X_{1}, X_{2} \in G_{TM}$ be related 
by a translation $T^{n}$.
We group successive symbols of $X_{1}$ and $X_{2}$ into $s_{0}$ and
$s_{1}$ configurations, then we group symbols of $Y_{X_{1}}$ and $Y_{X_{2}}$.
We repeat grouping until, say in the i-th grouping, we get for both 
$X_{1}$ and $X_{2}$ the local configurations of the same type, either 
$s_{0}$ or $s_{1}$, with supports containing both the origin
and $n$ (they are related by the same translation
as $X_{1}$ and $X_{2}$). It follows that  $C'_{X_{1}}(j)=C'_{X_{2}}(j)$,
for every $j>i$ and therefore $C'_{X_{1}}=C'_{X_{2}}+U$, where $U(j)=0$
for $j>i$.

Conversely, let $C'_{X_{1}}=C'_{X_{2}}+U$, where $U(j)=0$ for $j>i$.
Let $c(k)$ be a local configuration obtained by successive applications of
substitutions corresponding to $C'_{X_{k}}(l), l \leq i, k=1,2.$ 
If $c(1)$ and $c(2)$ are related by a translation, then 
$X_{1}$ and $X_{2}$ are related by the same translation. If $c(1)$ and $c(2)$
are related by a translation and the interchanging of 0's and 1's, 
then $X_{1}$ and $X_{2}$ are
related by the same translation and the interchanging of 0's and 1's. 
Observe that 
the first situation occurs if the absolute value of the difference
of the number of 1's in $C'_{X_{1}}(l)$ and $C'_{X_{2}}(l)$, $l \leq i$,
is even, otherwise we have the second case.
\vspace{1mm}
	     
\noindent {\bf b)} $X_{1}, X_{2} \in G_{TM}$ are related by the interchanging
of $0's$ and $1's$ if and only if $C'_{X_{1}}=C'_{X_{2}}+(10000...)$.
\vspace{1mm}

\noindent {\bf c)} $X_{1}, X_{2} \in G_{TM}$ are related
by the reflexion about
the origin if and only if $C'_{X_{1}}=C'_{X_{2}}+ (01111...)$.
\vspace{3mm}

Let us notice that the above group acts on the whole set $G_{TM}$ of
the Thue-Morse configurations. We will now construct a symmetry group
which leaves every Thue-Morse configuration invariant. This 
corresponds to crystal symmetries mentioned in the very beginning 
of the paper.

\noindent {\bf A symmetry group of the Thue-Morse configurations.}
\vspace{3mm}

\noindent The following theorem gives us positions of particles, i.e.,
$1'$s in a Thue-Morse configuration.
\begin{theo}
Let $X$ be a Thue-Morse configuration, with $1$ at the origin, 
and represented by $C'_{X} \in \cal W$. We have
\begin{equation}
X(i)=1 \; \; iff \; \; i=\sum_{k=0}^{\infty}(-1)^{(1+C'_{X}(k))}2^{k}a_{k},
\end{equation}
where $a_{k} \in \{0,1\}, k \geq 0$ and in the sum there are finite
and even number of nonzero terms.
\end{theo}
{\bf Proof:} Let us recall that the 
effect of performing the substitution to the right or to the left 
is the same as taking sequence
of symbols already obtained, changing every $0$ to $1$ and every $1$ to
$0$ and then placing the new sequence right to or left to the previous
sequence. We call such sequences blocks. If we start with $1$ at the origin, 
then every second block placed right to or left to the previous block 
begins with $1$. Let now $X(i)=1$. We identify the first block containing
the origin to which $i$ belongs and call it the exterior block. 
If $i$ is not at the beginning of this block, then we 
have to find out to which one of the two subblocks 
it belongs. We continue this procedure untill we identify the block 
such that $i$ is at the beginning of it. If the exterior block begins with
$1$, then its position is given by the sum in (1) with an even number
of nonzero terms and then we have to identify an even number 
of additional subblocks; to obtain (1) we have to add 
to the position of the left end of the exterior block 
an even number of different powers of $2$. 
In the other case, the exterior block begins with $0$ and its
position is given by the sum in (1) with an odd number of nonzero
$a_{i}$'s but we have to identify an odd number of additional subblocks.  
For example, in the right one-sided Thue-Morse sequence represented by 
$(1111111...)$, all exterior blocks begin with $1$ 
at the origin. $\Box$
\vspace{3mm}

Denote by $S_{TM}$ the set of all sequences
$\{a_{i},i \geq 0\}$
such that $a_{i}=1$ for finite and even number of i's and otherwise
$a_{i}=0$. $S_{TM}$ is a subgroup of $\otimes_{i=0}^{\infty}Z_{2}$.
It is generated by an infinite number of elements (with two successive
$1's$ and $0's$ otherwise). A Thue-Morse quasicrystal is generated
by successive applications of elements of $S_{TM}$ 
to the particle at the origin.

More precisely, positions of particles in a two-sided 
Thue-Morse configuration $X \in G_{TM}$ is a subset 
${\bf Z}_{X} \subset {\bf Z}$. The coding of $X$, $C'_{X}$, gives us
an addition, $+_{X}$, in ${\bf Z}_{X}$, such that $({\bf Z}_{X},+_{X})$
is a group. 

For example, ${\bf Z}_{X_{TM}}$ is generated by
$\{(-1)^{k+1}2^{k}:k \geq 0\}$.    
\vspace{3mm}

\noindent {\bf Acknowledgments}. Part of this work was done when
J. M. visited Laboratoire de Physique Th\'{e}orique et Math\'{e}matique,
Universit\'{e} Paris 7. J. M. thanks Laboratoire for hospitality
and a financial support and the Polish Scientific Committee for Research, 
for a financial support under the grant KBN 2 P03A 015 11. J. M. thanks
Aernout van Enter for many discussions and collaborations
on Thue-Morse systems and comments about this paper.

\end{document}